\documentclass[aps,prb,twocolumn,showpacs,showkeys,superscriptaddress,amsfonts,amsmath,amssymb, floatfix]{revtex4-1}
\usepackage[T1]{fontenc}
\usepackage[utf8]{inputenc}
\usepackage[english]{babel}
\usepackage{amsmath}
\usepackage{amsfonts}
\usepackage{amssymb}
\usepackage{array}
\usepackage{makeidx}
\usepackage{filecontents}
\usepackage{rotating}
\usepackage{graphicx}	
\usepackage{color}
\usepackage[left=2cm,right=2cm,top=2cm,bottom=2cm]{geometry}
\usepackage{wasysym}

\usepackage{dcolumn}
\usepackage{revsymb}
\usepackage{float}

\usepackage{inputenc}
\usepackage{ulem}
\usepackage{setspace}
\usepackage{multirow}
\usepackage{array}
\usepackage{tabularx}
\usepackage{bm}

\usepackage[colorlinks=true,linkcolor=blue,citecolor=blue,urlcolor=blue]{hyperref}

\begin{document}

\title{Infrared phonon spectroscopy on the Cairo pentagonal antiferromagnet Bi$_2$Fe$_4$O$_9$: a study through the pressure induced structural transition }
\author{M. Verseils}
\affiliation{Ligne AILES - Synchrotron SOLEIL, 91190 Gif-sur-Yvette CEDEX, France}

\author{A.P. Litvinchuk}
\affiliation{Texas Center for Superconductivity, University of Houston, Houston, TX 77204, USA}

\author{J-B. Brubach}
\author{P. Roy}
\affiliation{Ligne AILES - Synchrotron SOLEIL, 91190 Gif-sur-Yvette CEDEX, France}

\author{K. Beauvois}
\affiliation{Univ. Grenoble Alpes, CEA, IRIG, MEM, MDN, 38000 Grenoble, France and Institut Laue-Langevin, 38000 Grenoble, France}
\author{E. Ressouche}
\affiliation{INAC/MEM, CEA-Grenoble, 38042 Grenoble, France}

\author{V. Skumryev}
\affiliation{Institucio$'$ Catalana de Recerca i Estudis Avançats (ICREA), 08010 Barcelona, Spain and Departament de Fisica, Universitat Autonoma de Barcelona, 08193 Bellaterra, Spain}

\author{M. Gospodinov}
\affiliation{Institute of Solid State Physics, Bulgarian Academy of Sciences, 1784 Sofia, Bulgaria}

\author{V. Simonet}
\author{S. de Brion}
\affiliation{Universit\'e Grenoble Alpes, CNRS, Institut N\'eel, 38000 Grenoble, France}

\begin{abstract}
Magnetic and crystallographic transitions in the Cairo pentagonal magnet Bi$_2$Fe$_4$O$_9$ are investigated by means of infrared synchrotron-based 
spectroscopy as a function of temperature (20 - 300 K) and pressure (0 - 15.5 GPa). One of the phonon modes is shown to exhibit an anomalous softening as a function of temperature in the antiferromagnetic phase below 240 K, highlighting spin-lattice coupling. Moreover, under applied pressure 
at 40 K, an even larger softening is observed through the pressure induced structural transition. Lattice dynamical calculations reveal that this mode is indeed very peculiar as it involves a minimal bending of the strongest superexchange path in the pentagonal planes, as well as a decrease of the distances between second neighbor irons. The latter confirms the hypothesis made by Friedrich \textit{et al.,} \cite{friedrich_high-pressure_2012} about an increase in the oxygen coordination of irons being at the origin of the pressure-induced structural transition. As a consequence, 
one expects a new magnetic superexchange path that may alter the magnetic 
structure under pressure.
\end{abstract}

\maketitle

\section{INTRODUCTION}

\begin{figure*}
\centering
\includegraphics[width=0.9\textwidth]{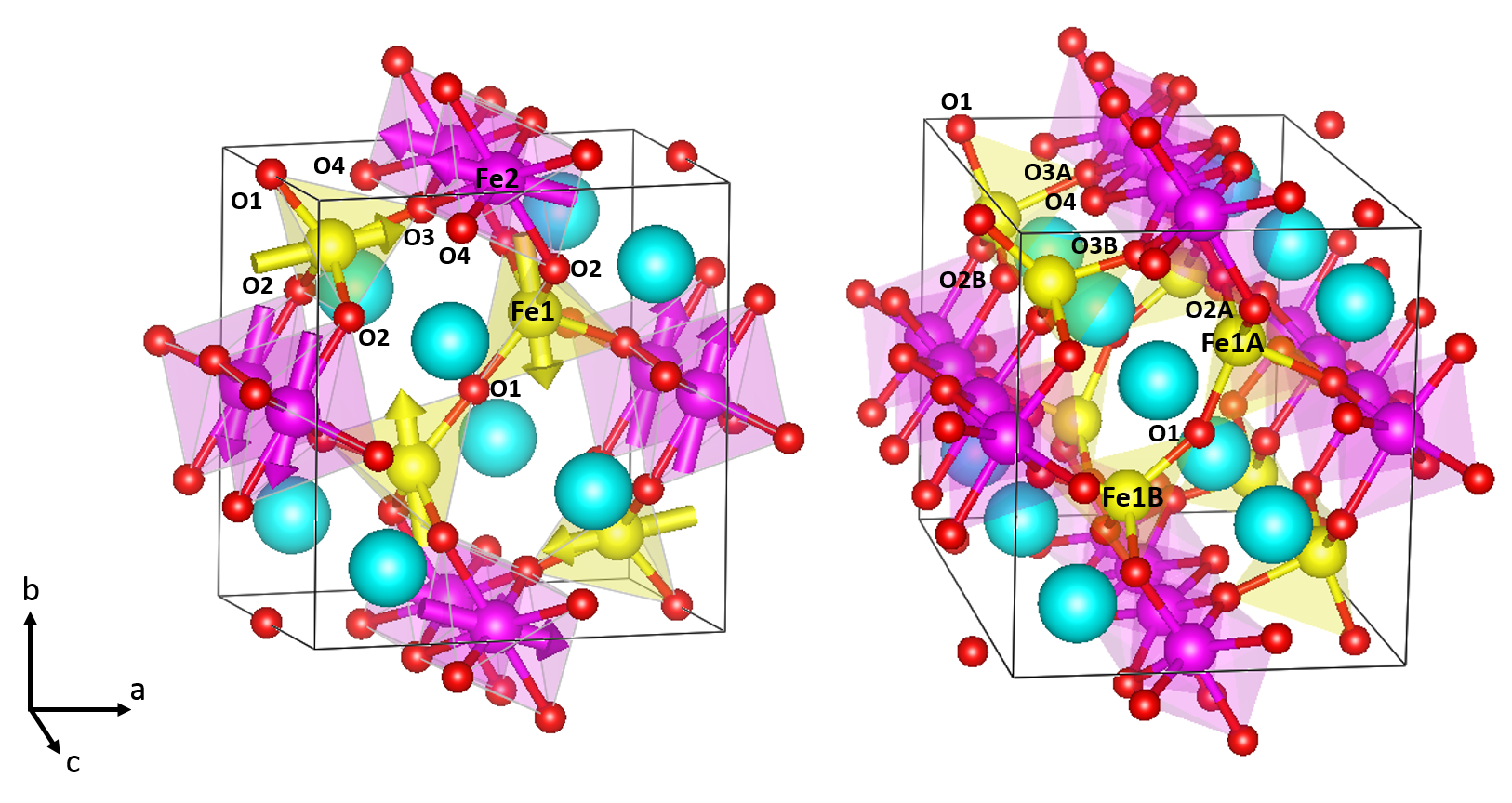}
\caption{(Color online) Left panel: The orthorhombic $Pbam$ unit cell of Bi$_2$Fe$_4$O$_9$ at low temperature and ambient pressure conditions. The 
arrows represent the spins orientation in the magnetic ordered phase (T < 
T$_N$) \cite{ressouche_magnetic_2009}. Right Panel: The orthorhombic $Pbnm$ unit cell of Bi$_2$Fe$_4$O$_9$ at high pressure (P > 7 GPa) and room temperature. \cite{friedrich_high-pressure_2012}. Blue: Bismuth atoms; Yellow: tetrahedral Irons (Fe1); Pink: Octahedral Irons (Fe2); Red: Oxygens. 
}
\label{Structure}
\end{figure*}

The Cairo pentagonal lattice is an original network containing irregular pentagons that are connected by their edges via three-fold and four-fold connected sites (see Figure \ref{Structure}). It has attracted interest lately because the pentagon building block has an odd number of bonds which can promote magnetic frustration as in the intensively studied triangle-based networks \cite{Gardner1999}. In such frustrated magnets, it has been predicted and observed experimentally that the stabilisation of a long 
range magnetic order may be impeached in favor of fluctuating states down 
to the lowest temperature, an archetypical example being the spin ice state on the pyrochlore lattice \cite{bramwell_spin_2001}. When the frustration can be partly released, thanks to other degrees of freedom such as lattice distortion for instance, complex magnetic order (non-collinear) may 
prevail \cite{blake_spin_2005,chapon_structural_2004}. These complex magnetic phases can furthermore induce a ferroelectric order when they break the centrosymmetry: these are type II multiferroics that have attracted considerable interest over the past 15 years \cite{kimura_magnetic_2003,lottermother_magnetic_2004,mostovoy_ferroelectricity_2006,wang_multiferroicity,sergienko_role_2006,sergienko_ferroelectricity_2006,khomskii_classifying_2009}.

In this article, we focus on one of the few experimental realizations of a Cairo pentagonal lattice, the bismuth iron oxide Bi$_2$Fe$_4$O$_9$ \cite{ressouche_magnetic_2009}. At ambient conditions, this compound crystallizes in the orthorhombic space group $Pbam$, No. 55, with two formula units per unit cell. This unit cell (see Figure \ref{Structure}) contains two distinct sites, Fe1 and Fe2, occupied by four Fe$^{3+}$ ions each, which have different oxygen coordinations (tetrahedral for Fe1 and octahedral 
for Fe2). The whole Fe network forms an analogue of the Cairo pentagonal lattice in the \textit{ab} plane with the noticeable difference that the fourfold connected site in the perfect lattice is replaced by a pair of Fe2 ions stacked along the \textit{c}-axis and sandwiching the planes containing the Fe1 ions (see left panel of Figure \ref{Structure}). The compound orders antiferromagnetically below T$_N$ = 238 K  in a complex non-collinear magnetic structure formed by two interlocked sublattices \textbf{associated with} the two different Fe1 and Fe2 sites where the spins are at 90$^\circ$ with respect to each other (see Figure \ref{Structure}) \cite{ressouche_magnetic_2009}. This remarkable magnetic arrangement is a direct signature of competing interactions and complex connectivity, which has been confirmed by theoretical studies \cite{rousochatzakis_quantum_2012,raman_su2-invariant_2005}. In Bi$_2$Fe$_4$O$_9$, the magnetic structure remains centrosymmetric and no associated ferroelectric order is expected although it has been reported for polycrystalline samples or nanoparticles \cite{singh_substantial_2008,tian_size_2009}.

Another interesting result concerning this material is that an isostatic pressure alters the crystallographic arrangement: above 7 GPa, Friedrich et al \cite{friedrich_high-pressure_2012} have reported a pressure induced structural transition at room temperature from $Pbam$ to $Pbnm$ (No. 62) centrosymmetric space-groups with the doubling of the unit cell along the $c$-axis ($c' = 2c$). The driving force of the transition was proposed to arise from the tendency of tetrahedral Fe1 to increase their oxygen 
coordination to five under pressure. Another clear signature of the transition is the large displacement of the O1 oxygen atoms that connect two Fe1 tetrahedra \cite{friedrich_high-pressure_2012} (see Figure \ref{Structure}). This pressure-induced structural transition raises the issue of its influence on the magnetic order observed at ambient pressure below 238 K, since it might strongly affect the magnetic super exchange paths enabling this order. 

To gain a better insight into this material, we have investigated the temperature and pressure evolution of the phonon modes that involve atomic displacements within the pentagonal plane by means of infrared measurements. For this purpose, we have used the intense infrared source available on the  AILES-B beamline at synchrotron SOLEIL combined with two different 
set-ups, one adapted to bulk sample allowing low-temperature measurements 
at ambient pressure and the second for high pressure and low temperature measurements adapted to sub-mm crystals \cite{voute_new_2016}. We have been able to identify all the B$_{2u}$ infrared phonon modes and to follow their evolution through the structural and magnetic phase transitions. \textbf{We evidenced magnetoelastic effects comparable to what is usually observed in antiferromagnets \cite{lockwood_spin-phonon_1988,kuzmenko_infrared_2001,schleck_infrared_2010,schleck_elastic_2010,lobo_infrared_2007}.} Finally, we show that at least one mode, around 225 cm$^{-1}$, has an abnormal behavior both as a function of temperature and pressure, whose implication on the magnetic properties is discussed.

\section{EXPERIMENTAL}

Single crystals of Bi$_2$Fe$_4$O$_9$ were synthesized using the high temperature solution growth method as described elsewhere \cite{iliev_phonon_2010}. Crystals were then preoriented using a Rigaku Xcalibur S 4-circles 
XRD diffractometer.

The infrared spectroscopy measurements were performed in the reflectivity 
configuration on the IFS125MR Michelson interferometer installed on the AILES beamline at SOLEIL.  A 6 $\mu$m Mylar beamsplitter and a 4.2 K bolometer were used to perform measurement in the far infrared (FIR) (60 - 850 
cm$^{-1}$) range with the resolution of 2 cm$^{-1}$. Measurements at ambient pressure on bulk sample were performed between room temperature and 20 K using a helium close-cycle cryostat. The optical setup allows a strong focusing and a quasi-normal incidence of the synchrotron beam onto the sample surface. The FIR waves were linearly polarized thanks to a polyethylene polarizer. The absolute reflectivity of the sample was obtained by using as reference the same gold-coated sample obtained by \textit{in situ} gold coating evaporation technique \cite{Homes_goldoverfilling_1993}.\\
High pressure measurements were realized at 40 K in the quasi-normal reflectivity geometry inside a diamond anvil cell (DAC) plugged to a cryostat 
into a high-pressure/low-temperature box \cite{voute_new_2016}. The diameter of the culets of the diamond anvil was 500 $\mu$m. A stainless steel gasket prindented and then drilled allowed to get a 250 $\mu$m large and 50 $\mu$m thick hole adapted to the sample size. This latter was preoriented and polished to be 30 $\mu$m thick with a \textit{Leica} mechanical polisher. Polyethylene powder was used as transmitting medium to fill the hole and a ruby ball was placed next to the sample to allow \textit{in situ} determination of the pressure using fluorescence \cite{}. The reflectivity of a gold foil in place of the sample at room temperature was used as the reference.

\section{RESULTS}

{\renewcommand{\arraystretch}{0.5}
\begin{table}[htbp!]
\center
\begin{turn}{0}
\begin{tabular}{lccc}
\hline
\hline
\vspace{0.01cm}\\
\textbf{Modes} & \textbf{Exp} & \textbf{LDC}  & \textbf{Main atomic motions} \\
B$_{2u}$& {\small cm$^{-1}$} & {\small cm$^{-1}$} &  \\
\hline
\vspace{0.01cm}\\
{\small (1)} &{\small  69} & {\small 93} & {\footnotesize Bi($y$)}   \\
{\small (2)}& {\small 110}  & {\small  139}& {\footnotesize Fe1($xy$)+O1($xy$)}  \\
{\small (3)} &{\small  128 }& {\small 180} & {\footnotesize Fe1($y$)+Fe2($y$)+O1($y$)+O2($xy$) + O3($y$)} \\
{\small (4)} & {\small 225} & {\small  202} & {\footnotesize Fe1($y$)+Fe2($y$)+O1($y$)+O2($y$)+O3($y$)}  \\
{\small (5)} &{\small  238}  & {\small 290} & {\footnotesize Fe1($x$)+Fe2($x$)+O1($xy$)+O3($y$)+O4($x$)} \\
{\small (6)} &{\small  310 }& {\small 325} & {\footnotesize Fe1($x$)+O1($xy$)+O3($xy$)}  \\
{\small (7)} & {\small 351} & {\small 383} & {\footnotesize O2($x$)+O3($x$)+O4($xy$)}  \\
{\small (8)} & {\small 399} &   {\small 409} & {\footnotesize O1($xy$)+O2($y$)+O3($x$)} \\
{\small (9)} & {\footnotesize $\sim$}\hspace{0.07cm}{\small 430} &  {\small 450} & {\footnotesize O1($x$)+O2($x$)+O3($y$)+O4($xy$)} \\
{\small (10)} & {\footnotesize $\sim$}\hspace{0.07cm}{\small 455} & {\small  499} & {\footnotesize Fe2($x$)+O2($x$)+O3($y$)+O4($x$)} \\
{\small (11)} & {\footnotesize $\sim$}\hspace{0.07cm}{\small 490} & {\small 613} & {\footnotesize O1($x$)+O2($xyz$)+O3($xy$)+O4($x$)} \\
{\small (12)} & {\small 605} &  {\small 639} & {\footnotesize O1($x$)+O3($y$)+O4($y$)} \\
{\small (13)} & {\small 645} &  {\small 789} & {\footnotesize O2($yz$)}  \\
{\small (14)} & {\small 805}  &  {\small 976} & {\small  O1($xy$)}\\
 &  & \\
\hline
\hline
\end{tabular}
\end{turn}
\caption{Frequencies of the B$_{2u}$ infrared active modes of Bi$_2$Fe$_4$O$_{9}$ determined experimentally at 300 K and calculated. The corresponding main atomic motions are also reported.}
\label{Assignation}
\end{table}
}

\subsection{$\Gamma$-point IR phonons}

\begin{figure*}
\centering
\includegraphics[width=0.9\textwidth]{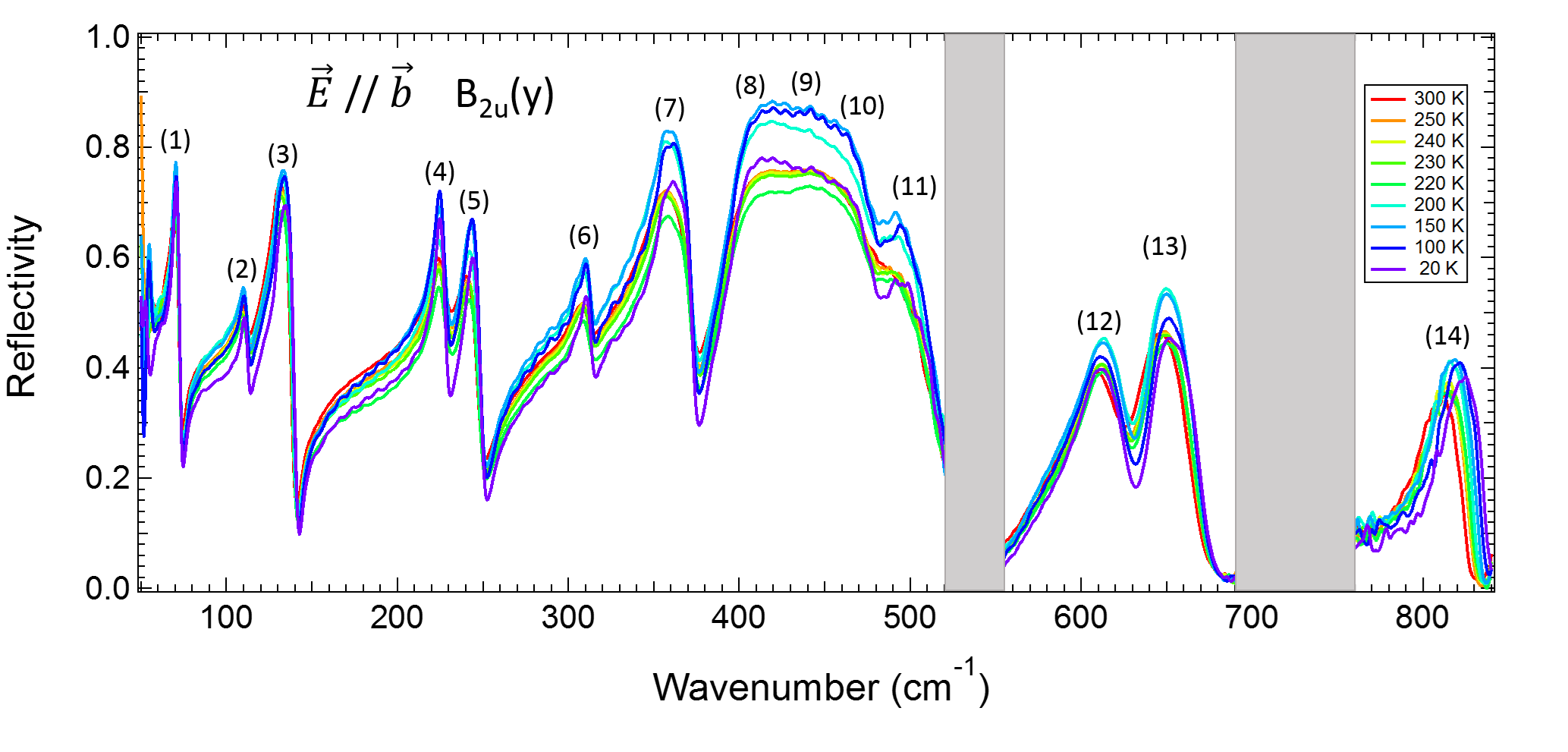}
\caption{Far-infrared reflectivity at quasi normal incidence of Bi$_2$Fe$_4$O$_9$ obtained at different temperatures, ambient pressure and with the electric field \textbf{E} of the electromagnetic wave along the $b$-axis of the sample. The two greyed out regions are not detectable because of 
destructive interferences of multiple reflections inside the beamsplitter. The numbers indicate the 14 B$_{2u}$ infrared active modes detected as expected from group theory in this configuration.}
\label{R_300K}
\end{figure*}

Group theory predicts $\Gamma_{Raman}$= 12 A$_g$ + 12 B$_{1g}$ + 9 B$_{2g}$ + 9 B$_{3g}$ = 42 Raman-active modes, $\Gamma_{IR}$ = 8 B$_{1u}$ 
+ 14 B$_ {2u}$ + 14 B$_{3u}$ = 36 infrared active modes and 9 A$_u$ silent modes in the low-pressure $Pbam$ space-group and 25 A$_g$ + 20 B$_{1g}$ + 25 B$_{2g}$ + 20 B$_{3g}$ = 90 Raman-active modes, 24 B$_{1u}$ + 19 B$_{2u}$ + 24 B$_{3u}$ = 67 infrared active modes and 20 A$_u$ silents mode in the high pressure \textit{Pbnm} space-group. At ambient pressure, for a FIR polarization along the b-axis, a total of 14 infrared active 
B$_{2u}$ modes are expected. The reflectivity spectrum obtained at ambient conditions with FIR electric field along the $b$-axis is presented in Figure \ref{R_300K}. Considering that the broad band between 400 and 500 cm$^{-1}$ is composed of 4 unresolved modes, we are able to identify all 14 B$_{2u}$ modes numbered in Fig \ref{R_300K}.

\begin{figure}[htbp!]
\centering
\includegraphics[width=0.9\columnwidth]{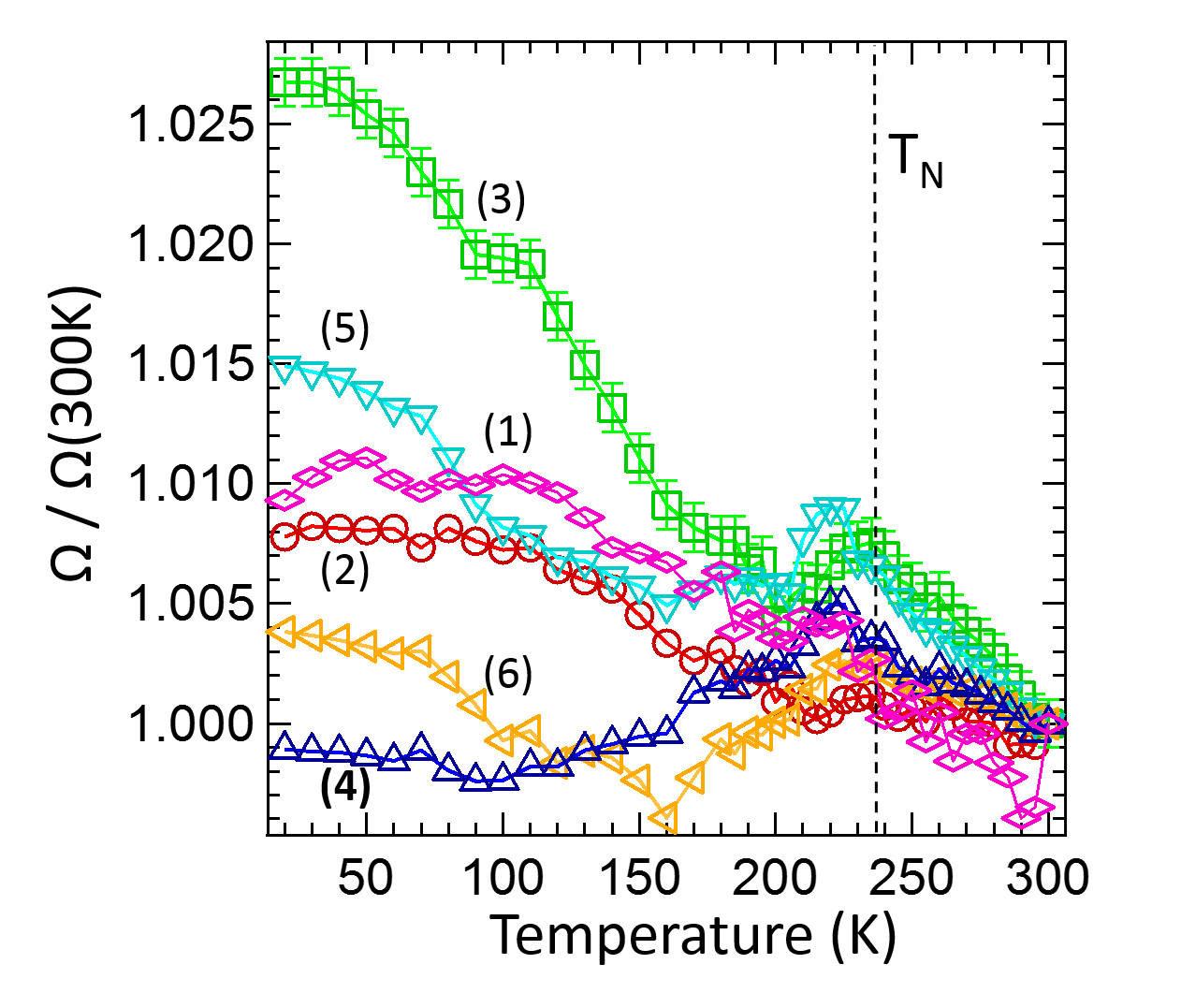}
\caption{Temperature dependence of normalized frequencies ($\Omega_k$) of 
modes B$_{2u}$(1), B$_{2u}$(2), B$_{2u}$(3), B$_{2u}$(4), B$_{2u}$(5) and B$_{2u}$(6) of Bi$_2$Fe$_4$O$_9$ from room temperature to 20 K. The reference is the frequency value at 300 K. The dashed line represents the Néel temperature of the sample, T$_N$ = 238 K. For the sake of clarity, the error bar is given only for mode (3) and was estimated from the reproducibility 
of the fits  by changing the number of modes in the model.}
\label{W_vs_Temp}
\end{figure}

Lattice dynamical calculations (LDC) for Bi$_2$Fe$_4$O$_9$ were performed 
within shell model using the general utility lattice program (GULP) \cite{gale_gulp_1997}, which is known to adequately describe phonon modes of oxides with various crystallographic structures \cite{Popov_1995,Litvinchuk_2004,PhysRevB.75.104118,Litvinchuk_2009}. In the shell model, each ion is considered to be a point core surrounded by a massless shell. The free 
ion polarizability is accounted for by a force constant. The short range potentials V(r) are chosen in the Born–Mayer–Buckingham form as follows:$$V(r) = a\exp(-r/r_0) -cr^6$$ where $r$ is the interatomic distance. The Coulomb energy calculations are based on a real space summation involving a spherical cut-off boundary, which is defined by the cut-off radius (set to 12 \AA), which makes the sum of all charges within 
the spherical cut-off region equal to zero. In our calculations, we used the same set of shell model parameters and short range potentials, which provides frequencies closest to the experimental Bi$_2$Fe$_4$O$_9$ Raman scattering study  \cite{iliev_phonon_2010}. The calculated frequencies and the main atomic motions involved in the infrared modes are given in Table \ref{Assignation}. All experimental frequencies, except the ones of modes (9), (10) and (11) were obtained by fitting the data with RefFit software \cite{RefFit} using the Drude Lorentz (DL) model for the dielectric function of insulating materials, as described in the supplementary information. We fitted the data under the constraint that the reflectivity calculated at normal incidence at the air/sample interface, $R = |1-\sqrt{\epsilon}|^2/|1+\sqrt{\epsilon}|^2$, matches the experimental value.

The assignment proposed in Table \ref{Assignation} allows to identify all 
the 14 experimental modes. The modes \textbf{associated with} heavier atoms are located at the lower frequencies: the first mode in particular is attributed mainly to the Bi atom displacements, modes (2) to (6) to Fe and oxygens displacements, while displacements of the oxygens alone are involved in modes (7) to (14). The mean deviation between experimental and calculated phonon frequencies is found to be 10\%, which might be considered as reasonable, remembering that parametric shell model does not take into account electronic correlations. Note that our measurements are in qualitative agreement with previous infrared studies on Bi$_2$Fe$_4$O$_9$ \cite{voll_variation_2006,debnath_series_2010} that do not report any experimental modes above 850 cm$^{-1}$ either.

\subsection{Temperature dependence of IR phonons}

Reflectivity measurement of Bi$_2$Fe$_4$O$_9$ between 20 and 300 K were performed at quasi-normal incidence with the electric field along the $b$-axis of the bulk sample. The reference for each spectrum is the reflectivity of the sample covered with gold measured at the same temperatures. As 
already known from previous studies \cite{shamir_magnetic_nodate,ressouche_magnetic_2009}, when lowering temperature, Bi$_2$Fe$_4$O$_9$ undergoes an antiferromagnetic transition at $T_N$ = 238 K while no structural transition has been reported. As expected, our measurements reported in Figure \ref{R_300K} show that the number of modes (14 B$_{2u}$) is maintained through the whole temperature range. Here again, we used the DL model described previously, and fitted the spectra at each temperature using the 
RefFit software \cite{RefFit}. The modes between 50 and 250 cm$^{-1}$ can 
be fitted with high accuracy and small error bars at each temperature as they are narrow and well separated.

 \textbf{In Figure \ref{W_vs_Temp}, the normalized frequencies of modes (1)-(6) obtained from the DL fits are displayed as a function of temperature. We also report all the fit parameters in the appendix.} The conventional temperature dependence of phonon \textbf{modes is described by the well known relations, reported by Balkanski \cite{balkanski_anharmonic_1983}: due to anharmonicity of the potential, a progressive frequency hardening is expected with decreasing temperature, which levels off at low temperature. With the exception of mode (1) and (2), all modes show a pronounced deviation from the Balkanski's behavior, characterized by an abnormal softening 
starting just below T$_N$.} The softening stops at 200 K for mode (3), and at 160 K for modes (5) and (6). At lower temperature, all these modes recover the conventional hardening. Noticeably, mode (4) starts to soften at T$_N$ and never recovers the Balkanski's behavior at low temperature. These different softenings suggest that the antiferromagnetic interactions responsible for the long-range antiferromagnetic order below T$_N$ affect the atomic displacements involved in modes (3) to (6) through spin-lattice coupling. This effect is maximal for mode (4) that displays a singular behavior with the strongest deviation from conventional phonon thermal 
evolution.

\subsection{IR phonons under high pressure}

\begin{figure*}[htbp!]
\centering
\includegraphics[width=0.9\textwidth]{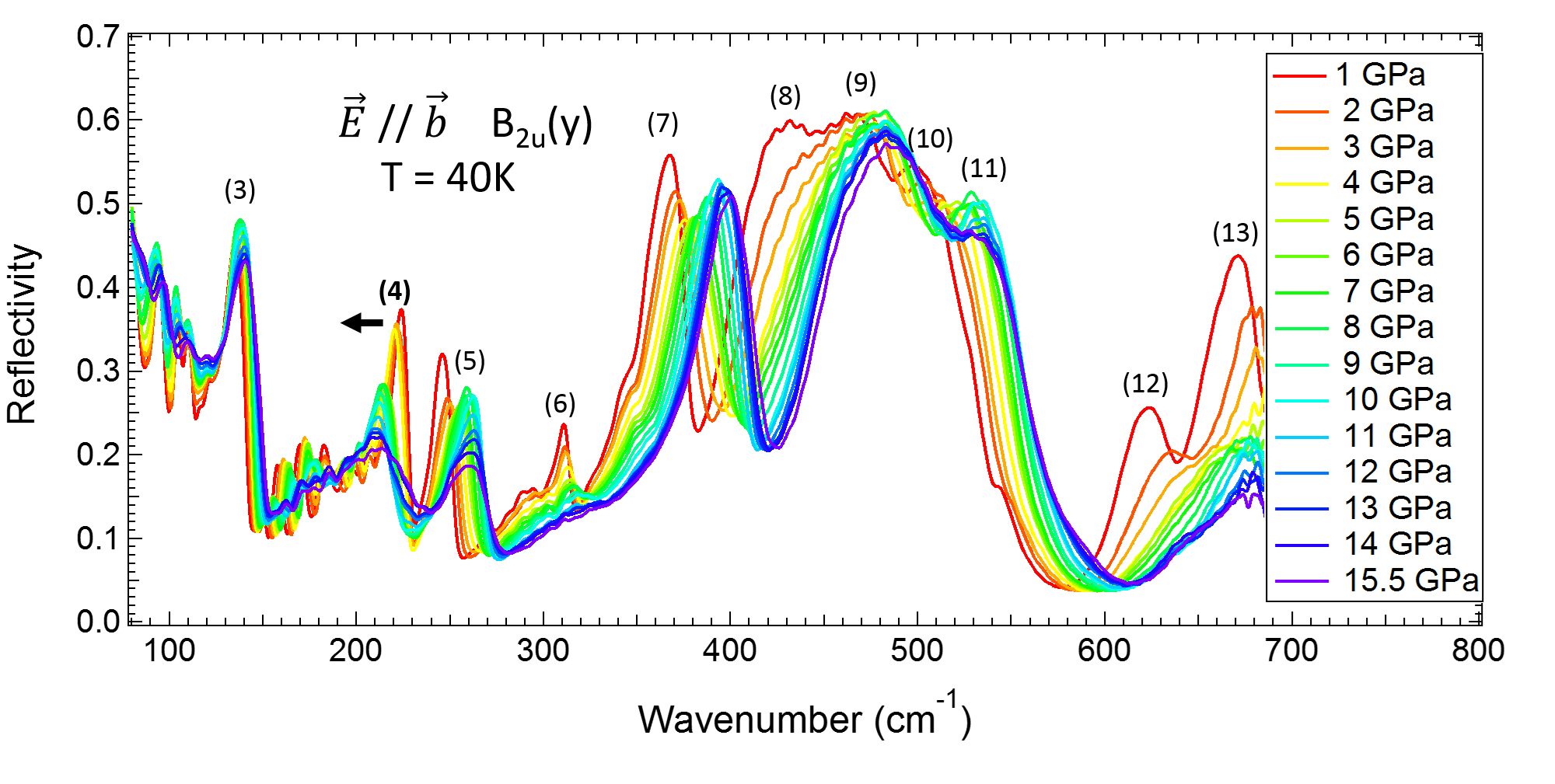}
\caption{Far-infrared reflectivity at quasi normal incidence of a small sample ($\sim$ 200 $\mu$m) of Bi$_2$Fe$_4$O$_9$ obtained in the diamond anvil cell and with electric field E of the electromagnetic wave along the b-axis. The reference used is the reflectivity of a piece of gold leaf of 
approximately the same size than the sample and placed in the hole of the 
gasket of the DAC. The numbers indicate the B$_{2u}$ infrared active modes already identified in the bulk sample.}
\label{Refl_press}
\end{figure*}

In order to complete our study and to investigate the pressure induced structural transition reported by Friedrich et al., \cite{friedrich_high-pressure_2012} we measured the reflectivity of Bi$_2$Fe$_4$O$_9$ from 1 to 15.5 GPa with the electric field of the electromagnetic wave along the $b$-axis at a temperature of 40 K (well below the reported antiferromagnetic transition at ambient pressure). These spectra are displayed in Figure \ref{Refl_press}. They are in good agreement with the spectrum of the bulk measured as a function of temperature. Due to diffraction, multiple reflections inside the diamond anvil cell and beamsplitter limitations, the low frequency (under 100 cm$^{-1}$) and the high frequency (above 700 cm$^{-1}$) parts of the spectra are not usable. Therefore modes (1), (2) and 
(14) detected in bulk sample are not visible in the high pressure spectra 
of figure \ref{Refl_press}. Nevertheless, all the other modes detected in 
bulk sample are recovered and are numbered as in figure \ref{R_300K}. Interestingly, in the 1 GPa spectrum, we can identify unambiguously all four 
modes of the broad band between 450 and 550 cm$^{-1}$, which confirms our 
previous hypothesis.

\begin{figure}
\centering
\includegraphics[width=0.9\columnwidth]{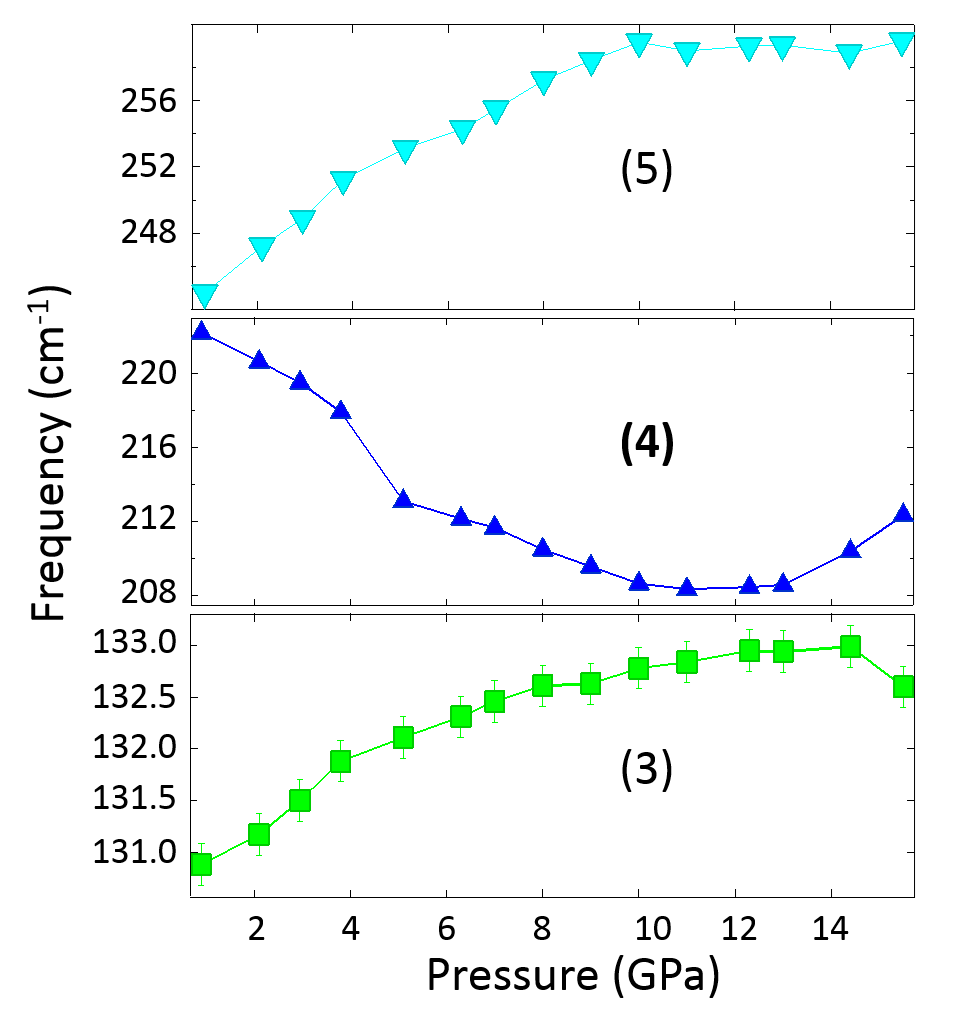}
\caption{ Pressure dependence of frequencies of modes (3), (4) and (5) obtained from Drude-Lorentz fits of 40 K measurements. The error bars have been estimated using the reproducibility of the fit with different numbers of oscillators. For modes (4) and (5), the errors bars are smaller than 
the symbols.}
\label{W_vs_P}
\end{figure}

According to Friedrich {\it et al.} \cite{friedrich_high-pressure_2012}, Bi$_2$Fe$_4$O$_9$ undergoes a structural transition from $Pbam$ toward $Pbnm$ with the doubling of the unit cell at about 7 GPa at ambient temperature. In the high pressure space-group $Pbnm$ (Z=4), group theory predicts 19 B$_{2u}$ infrared active modes. \textbf{ We expect to observe, apart from the initial 14 B$_{2u}$ singlet modes,  5 additional ones of the same symmetry}.  In contrast, our data show no additional modes when pressure is increased up to 15.5 GPa. \textbf{From structural refinements of X-ray data under pressure \cite{friedrich_high-pressure_2012}, it is established that the lattice distortion in the $ab$ plane is  quite small and probably too weak to induce new infrared detectable phonons. Moreover, the shape of a phonon band measured by reflectivity can be quite large because spreading from transverse (resonance) to longitudinal frequencies thus causing overlapping of close modes. Note that some new modes have been observed by Raman spectroscopy at the pressure-induced transition \cite{friedrich_high-pressure_2012}, but not all the predicted ones}. Our IR data evidence a substantial broadening of modes (4), (5) and (6) with pressure. All the IR modes, with the noticeable exception of mode (4), shift toward higher frequencies with increasing pressure, as expected for a normal lattice contraction and hardening of forces between atoms.

As described by Langerome {\it et al.} \cite{langerome_probing_2019}, the 
interface with a diamond affects the baseline in the low frequency side of the reflectivity band and should be considered in the fitting model of quasi-normal reflectivity. Using the RefFit software \cite{RefFit}, we attempted to simulate the effect of the interface with the diamond, but found out only a negligible effect of the diamond interface with respect to interferences and diffraction effects dominating at low frequencies. We thus have fitted the reflectivity data at each pressure of the low energy modes. The results \textbf{for the modes frequencies} are presented in Figure \ref{W_vs_P} for modes (3), (4) and (5). \textbf{All the fit parameters are reported in the appendix.} As mentioned before, modes (3) and (5) 
follow the same hardening behavior when pressure is increased while mode (4) undergoes instead a very pronounced softening of $\sim$ 14 cm$^{-1}$ between 1 and 10 GPa. Above 10 GPa, the pressure dependence evolution changes and the frequency slowly increases with pressure. Note that there is 
no clear discontinuity marking the pressure induced structural transition, which has however been confirmed to occur at low temperature by neutron 
diffraction, as well as the presence of magnetic order \cite{Beauvois}.

\begin{table*}
\center
\begin{tabular}{|c|c|c|c|c|}
\hline
\hline
 && J$_3$ & J$_4$ & J$_5$  \\
\hline
\multirow{3}{*}{Equilibrium values}  & $\alpha (^\circ)$ & 129.7 ($\times 
4$)&  180  &  119.6 ($\times 4$)    \\
&$d_{Fe1 -O}$ (\AA) & 1.847 ($\times 2$)   & 1.811 ($\times 2$) & 1.901 ($\times 4$) \\
&$d_{Fe2 -O}$ (\AA) & 1.972 ($\times 4$) &  & 2.029 ($\times 4$) \\
\hline
\multirow{7}{*}{Mode (1)}& $\Delta \alpha (^\circ)$&  -0.9 ($\times 4$)& -1.4  &  -0.1 ($\times 2$)\\
&$\Delta d_{Fe1 - O}$ (\AA)& -0.04 ($\times 2$)  &$\pm$ 0.001   &-0.003 ($\times 2$)   \\
&$\Delta d_{Fe2 - O}$ (\AA)& -0.038 ($\times 4$)  &  & -0.134 ($\times 2$)  \\
&&  &   \\
&$\Delta \alpha (^\circ)$&  &  & + 0.86 ($\times 2$)\\
&$\Delta d_{Fe1 - O}$ (\AA)&  &  & -0.018 ($\times 2$) \\
&$\Delta d_{Fe2 - O}$ (\AA)&  &  & +0.138 ($\times 2$) \\
\hline
\multirow{7}{*}{Mode (2)}&$\Delta \alpha (^\circ)$& +1.5 ($\times 2$) & \textbf{-11.7}  & +\textbf{7.93} ($\times 2$)  \\
&$\Delta d_{Fe1 - O}$ (\AA)& +0.117   & +0.014/+0.006  & +0.008 ($\times 2$) \\
&$\Delta d_{Fe2 - O}$ (\AA)& +0.081 ($\times 2$)  &  & +0.011 ($\times 2$)  \\
&&  &  &  \\
&$\Delta \alpha (^\circ)$& -1.5 ($\times 2$)  &  & +0.42 ($\times 2$) \\
&$\Delta d_{Fe1 - O}$ (\AA)& -0.114 &  & +0.141 ($\times 2$)  \\
&$\Delta d_{Fe2 - O}$ (\AA)& -0.072  ($\times 2$)&  & +0.018 ($\times 2$) 
\\
\hline
\multirow{7}{*}{Mode (3)}&$\Delta \alpha (^\circ)$ &+ 0.5 ($\times 2$) & \textbf{-10.7} & -0.32 ($\times 2$)  \\
&$\Delta d_{Fe1 - O}$ (\AA)& +0.18 & +0.014/+0.002  & +0.03 ($\times 2$)  
\\
&$\Delta d_{Fe2 - O}$ (\AA)&-0.034 ($\times 2$) &  & -0.136 ($\times 2$)  
\\
&&  &  & \\
&$\Delta \alpha (^\circ)$&  -1.9 ($\times 2$) &  & \textbf{+2.86} ($\times 2$) \\
&$\Delta d_{Fe1 - O}$ (\AA)&-0.068   &  & -0.027 ($\times 2$)  \\
&$\Delta d_{Fe2 - O}$ (\AA)& +0.044 ($\times 2$)&  & -0.139 ($\times 2$) \\
\hline
\multirow{7}{*}{Mode (4)}& $\Delta \alpha (^\circ)$ &\textbf{-3.8} ($\times 2$) & \textbf{-7.5} & \textbf{-3}  ($\times 2$) \\
&$\Delta d_{Fe1 - O}$ (\AA)&+0.179   & 0/+0.009  &-0.019 ($\times 2$)  \\
&$\Delta d_{Fe2 - O}$ (\AA)& -0.147 ($\times 2$) &  & \textbf{-0.485} ($\times 2$)  \\
&&  &  & \\
&$\Delta \alpha (^\circ)$&  \textbf{+2.4} ($\times 2$) &  & -0.3 ($\times 
2$) \\
&$\Delta d_{Fe1 - O}$ (\AA)&-0.172   &  & -0.021 ($\times 2$)  \\
&$\Delta d_{Fe2 - O}$ (\AA)& +0.162 ($\times 2$)&  & \textbf{-0.488} ($\times 2$) \\
\hline
\multirow{7}{*}{Mode (5)} & $\Delta \alpha (^\circ)$ &\textbf{-5.2} ($\times 4$) & \textbf{-28.3} & \textbf{-6.1} ($\times 2$)  \\
&$\Delta d_{Fe1 - O}$ (\AA)& \textbf{-0.415} ($\times 2$) & +0.025/+0.089 
 & +0.128 ($\times 2$)  \\
&$\Delta d_{Fe2 - O}$ (\AA)& +0.062 ($\times 4$)  &  &-0.129 ($\times 2$) 
  \\
&&  &  &  \\
&$\Delta \alpha (^\circ)$&  &  & +0.74 ($\times 2$) \\
&$\Delta d_{Fe1 - O}$ (\AA)&   &  & -0.048 ($\times 2$)  \\
&$\Delta d_{Fe2 - O}$ (\AA)& &  & -0.138 ($\times 2$) \\
\hline
\multirow{7}{*}{Mode (6)}& $\Delta \alpha (^\circ)$ &\textbf{-20.4} ($\times 2$) & \textbf{-52.8} & \textbf{-10.4}  ($\times 2$) \\
&$\Delta d_{Fe1 - O}$ (\AA)&+0.798  & 0.27/+0.152  &+0.136 ($\times 2$)  \\
&$\Delta d_{Fe2 - O}$ (\AA)& -0.093 ($\times 2$) &  & -0.153 ($\times 2$) 
 \\
&&  &  &  \\
&$\Delta \alpha (^\circ)$&  \textbf{+6.44} ($\times 2$) &  & \textbf{-2.23 }($\times 2$) \\
&$\Delta d_{Fe1 - O}$ (\AA)&\textbf{-0.707}   &  & \textbf{0.247} ($\times 2$) \\
&$\Delta d_{Fe2 - O}$ (\AA)& \textbf{+0.322} ($\times 2$)&  & -0.145 ($\times 2$) \\
\hline
\hline
\end{tabular}
\caption{Comparison between the superexchange paths in the pentagonal lattice (Fe1-O-Fe2 for the J3 and J5 interactions and Fe1-O-Fe1 for the J4 one) at equilibrium and ambient conditions and altered by the atomic vibration \textbf{associated with} the B2u modes. The degeneracy of each motion in terms of angle and Fe-O distances is given in brackets. The larger calculated changes are highlighted in bold.}
\label{Tableau2}
\end{table*}

\begin{table*}
\center
\begin{tabular}{|c|c|c|c|}
\hline
\hline
 & & J$_6$ & J$_6'$   \\
\hline
\multirow{4}{*}{Equilibrium} & $ d_{Fe1-Fe1}$ (\AA) & 4.241 & 4.531   \\
& $d_{Fe1-O3}$ (\AA) & 3.021& 3.641   \\
& $d_{O3-Fe1}$ (\AA) & 1.847&  1.847  \\
& $ \alpha$ ($^\circ$) & 119.2&   106.7 \\
\hline
\multirow{4}{*}{Mode (1)} & $\Delta d_{Fe1-Fe1}$ (\AA) & $+$0.004/$-$0.005 & $+$0.012/$-$0.012 \\
&$\Delta d_{Fe1-O3}$ (\AA) & $-$0.03/$+$0.027& $-$0.06/$+$0.059   \\
& $\Delta d_{O3-Fe1}$ (\AA) &$-$0.04/$+$0.042 & $+$0.042/$-$0.04   \\
& $\Delta \alpha$ ($^\circ$) &$+$3.1/$-$3 & $+$1.6/$-$1.7   \\
\hline
\multirow{4}{*}{Mode (2)} & $\Delta d_{Fe1-Fe1}$ (\AA) & $-$0.12/$+$0.146 
& $-$0.36/$+$0.361 \\
&$\Delta d_{Fe1-O3}$ (\AA) & $-$0.25/$+$0.259& $-$0.05/$+$0.286   \\
& $\Delta d_{O3-Fe1}$ (\AA) &$+$0.107/$-$0.11 & $-$0.08/$+$0.117   \\
& $\Delta \alpha$ ($^\circ$) &$+$1.1/$-$0.3 & $-$0.9/$+$0.9   \\
\hline
\multirow{4}{*}{Mode (3)} & $\Delta d_{Fe1-Fe1}$ (\AA) & $-$0.14/$+$0.194 
& $-$0.46/$+$0.465 \\
&$\Delta d_{Fe1-O3}$ (\AA) & 0.0/$-$0.002& $-$0.023/$+$0.022   \\
& $\Delta d_{O3-Fe1}$ (\AA) &$+$0.18/$-$0.07 & $-$0.07/$+$0.18   \\
& $\Delta \alpha$ ($^\circ$) &$-$12.4/$+$0.9 & $-$15.2/$+$13.4   \\
\hline
\multirow{4}{*}{Mode (4)} & $\Delta d_{Fe1-Fe1}$ (\AA) & \textbf{$-$0.16}/$+$0.231 & $-$0.53/$+$0.544 \\
&$\Delta d_{Fe1-O3}$ (\AA) & \textbf{$-$0.33}/$+$0.459& $-$0.73/$+$0.738  
 \\
& $\Delta d_{O3-Fe1}$ (\AA) &\textbf{$-$0.17}/$+$0.179 & $+$0.179/$-$0.17 
  \\
& $\Delta \alpha$ ($^\circ$) &\textbf{$+$17.4}/$-$13.6 & $+$0.1/$-$1.8   \\
\hline
\multirow{4}{*}{Mode (5)} & $\Delta d_{Fe1-Fe1}$ (\AA) & $-$0.021/$+$0.021 & $-$0.056/$+$0.057 \\
&$\Delta d_{Fe1-O3}$ (\AA) & $+$0.093/$-$0.02& $-$0.36/$+$0.315   \\
& $\Delta d_{O3-Fe1}$ (\AA) &$-$0.42/$+$0.424 & $+$0.456/$-$0.42   \\
& $\Delta \alpha$ ($^\circ$) &$+$13.6/$-$12.1 & $-$1.5/$+$0.5   \\
\hline
\multirow{4}{*}{Mode (6)} & $\Delta d_{Fe1-Fe1}$ (\AA) & $-$0.1/$+$0.111 & $-$0.3/$+$0.279 \\
&$\Delta d_{Fe1-O3}$ (\AA) & $-$0.88/$+$0.905& $-$0.32/$+$0.488   \\
& $\Delta d_{O3-Fe1}$ (\AA) &$+$0.798/$-$0.71 & $-$0.72/$+$0.798   \\
& $\Delta \alpha$ ($^\circ$) &$+$0.2/$-$14.8 & $+$32.8/$-$19.1   \\
\hline
\hline
\end{tabular}
\caption{ Comparison between the superexchange paths Fe1-O3-Fe1 involved in J$_6$ and J$_6'$ interactions at equilibrium and ambient conditions and altered by the atomic vibration \textbf{associated with} the B2u modes. 
The simultaneous decrease of the Fe1-Fe1 and Fe1-O3 bond length and increase of Fe1-O3-Fe1 angle of interaction J$_6$  are underlined in bold.}
\label{Tableau3}
\end{table*}

\section{Discussion}

We have shown that, at ambient pressure, due to magnetoelastic effect at the Néel temperature, phonon modes  (3), (4), (5) and (6) (reported in Table \ref{Assignation}) clearly exhibit a deviation from the conventional anharmonic behavior, like in other magnetoelastic materials \cite{cao_magnetoelastic_2008,cao_spin-lattice_2008}. These modes imply strong displacements of the Fe1 and Fe2 atoms, and of the oxygens (O1, O2 and O3) mediating the superexchange paths. For modes (3), (5) and (6), the abnormal softening stops at 160 K whereas mode (4) continues to soften until the lowest measured temperature. Besides, when the pressure is further increased in the magnetically ordered phase, one observes the expected hardening of all modes, except mode (4), which softens.

\begin{figure}[htbp!]
\centering
\includegraphics[width=0.8\columnwidth]{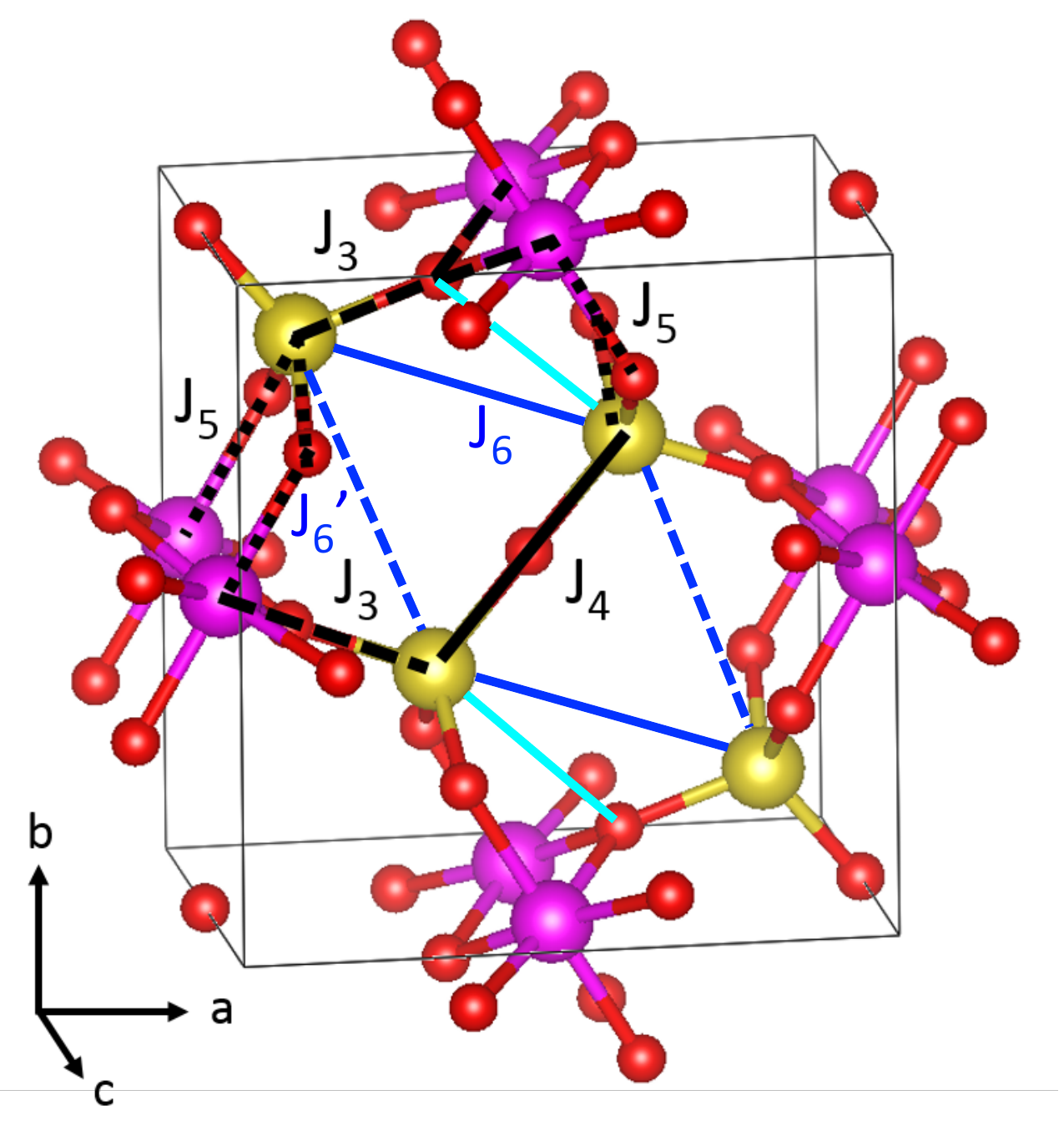}
\caption{Perspective view of a pentagonal slab of Bi$_2$Fe$_4$O$_9$. The Fe1 (in yellow) and the Fe2 (in pink) magnetically interact by the super-exchange interactions J$_3$ (dashed line), J$_4$ (line) and J$_5$ (dotted 
line). The second neighbor interactions between Fe1, J$_6$ and J$_6'$, are shown in blue, as well as the potential Fe1-O3 new atomic bond at high pressure in dotted turquoise. For sake of clarity the Bi atoms are absent 
in the drawing.}
\label{Interactions}
\end{figure}

In order to understand the distinctive feature of this mode and its relevance to the structural and magnetic properties, we calculated the dynamical changes, induced by each mode, in the bond super-exchange paths forming the pentagonal lattice. As reported in Figure \ref{Interactions}, within the pentagonal lattice, the iron atoms magnetically interact via 3 competing antiferromagnetic super-exchange integrals, J$_3$, J$_4$ and J$_5$ corresponding to the exchange paths Fe1-O3-Fe2, Fe1-O1-Fe1 and Fe1-O2-Fe2, respectively. Note that two additional rather weak superexchange interactions, J$_1$ and J$_2$, are at play between the Fe2 along the c-axis. According to the Goodenough-Kanamori-Anderson rule \cite{goodenough_theory_1955,Anderson1950}, the antiferromagnetic super-exchange interaction between two transition metals is the strongest when the angle is 180$^\circ$ and it weakens as the angle decreases. Moreover, the strength of the interaction increases when the bond length distances between the magnetic atom and the oxygen decreases. At the equilibrium, the values of the super-exchange interactions have been determined recently from inelastic neutron 
measurements \cite{beauvois_dimer_2020}: J$_1$ = 3.7, J$_2$ = 1.3, J$_3$ = 6.3,  J$_4$ = 24, and J$_5$ = 2.9 meV. The dominating antiferromagnetic interaction is J$_4$ as its exchange path angle is exactly 180$^{\circ}$ at ambient conditions.

The LDC calculations gives us the maximal atomic displacements for each phonon mode in the low-pressure \textit{Pbam} phase. We used these results 
to calculate the variation induced by each oscillation mode in the super-exchange paths \textbf{associated with} J$_3$, J$_4$ and J$_5$ (angle and 
bond length distances) and in the Fe1-O3-Fe1 paths possibly enabling the second neighbor interactions J$_6$ and J$_6'$ between the tetrahedral irons (see Figure \ref{Interactions}). The changes in the superexchange angle $\Delta \alpha$ and the bond length variations $\Delta d$ is given for each mode in Tables \ref{Tableau2} and \ref{Tableau3}. The corresponding oscillations may induce a splitting of the otherwise equivalent angle and 
distances of the equilibrium structure. All modes thus produce a disymetrization of the Fe1-O1-Fe1 J$_4$ exchange path and two inequivalent exchange paths for J$_5$. For J$_3$, only modes (2)-(4) and (6) lead to a path splitting. The distance and angle changes \textbf{associated with} mode (1) are small as expected since it reflects mainly the bismuth motion. All 
the other modes significantly modify some of the angles and distances involved in the exchange paths of J$_3$, J$_4$ and J$_5$. The most important 
change is the systematic reduction of the Fe1-O1-Fe1 angle entering the J$_4$ interaction. One major crystallographic change induced by pressure is actually the kinking of the straight Fe1-O1-Fe1 angle between the FeO$_4$ tetrahedra, which is incidentally expected to weaken the J$_4$ interaction. It is \textbf{associated with} the displacement of the O1 oxygen atom from its special position on an inversion centre to a less-constrained 
position on a mirror plane \cite{friedrich_high-pressure_2012}. In addition to the superexchange paths related to the J$_3$ to J$_5$ pentagonal interactions, the second neighbor Fe1-O3 distance (light blue line in Figure \ref{Interactions}) is also altered by the different phonon motions as can be seen in Table \ref{Tableau3}. A sufficient reduction of this bond length is expected to enable a new superexchange interaction J$_6$ through the second neighbor path Fe1-O3-Fe1. 

Let us now consider the distinctive character of mode (4). It involves both iron sites Fe1 and Fe2 and all oxygens atoms involved in the pentagonal exchange paths (Table \ref{Assignation}), and it displays the minimum change of the Fe1-O1-Fe1 180$^{\circ}$ angle, hence suggesting a minimum reduction of the dominant J$_4$ interaction (Table \ref{Tableau2}). Remarkably, mode (4) is also the one involving the strongest simultaneous reduction of both Fe1-Fe1 and Fe1-O3 bond lengths and increase of the Fe1-O3-Fe1 119 $^{\circ}$ angle involved in the J$_6$ exchange path, therefore strengthening it ((Table \ref{Tableau3}). In particular, the second neighbor Fe1-O3 bond length decreases from the equilibrium value of 3.02 to 2.69 
\AA, which may allow this O3 to enter in the coordination sphere of the Fe1 atom. Concerning the J$_6'$ interaction, the second neighbor Fe1-O3 bond length also decreases significantly but only to reach $\approx$2.9 \AA, which remains out of the coordination sphere of the Fe1. Note that mode 
(4) softening under pressure is as large as 7\%, while its softening below T$_N$ is only 0.5 \%. These softenings imply that the atomic motions of 
mode (4) are energetically favored by the application of pressure as well 
as by the magnetic ordering. Actually, according to Friedrich {\it et al} 
\cite{friedrich_high-pressure_2012}, the driving force of the structural transition, also observed in Bi$_2$Ga$_4$O$_9$ at higher pressure, is the 
tendency of Fe$^{3+}$ and Ga$^{3+}$ in tetrahedral coordination towards a 
higher coordination than four on pressure increase. For mode (6), the second neighbor Fe1-O3 bond length also decreases but this is concomitant to 
an increase of the first neighbor Fe1-O3 bond length entering the J$_6$ interaction. 
Thus, the dynamical increase of oxygen coordination is clearly a specificity of mode (4). On the magnetic side, it means that another exchange path (J$_6$) could be enabled between second Fe1 neighbors, which is expected to modify the magnetic order in the high pressure phase. Friedrich {\it 
et al.} have also pointed out the flexibility of the J$_4$ interactions under pressure that is confirmed by the strong decrease of the Fe1-O1-Fe1 angle in all modes. However, this change is minimal for mode (4) which might indicate that, in the magnetically ordered phase, the bending of the J$_4$ angle is opposed by magnetostrictive effects.

Globally, these results highlight the interplay between magnetism and atomic motions. As a consequence, the magnetic structure stabilized at ambient pressure below T$_N$ could be strongly modified in the pressure-induced phase. Indeed, the change in the J$_4$ interaction as well as the possibility of new exchange paths can release the frustration at the origin of 
the peculiar 90$^{\circ}$ magnetic structure. Modifications in the magnetic structure under pressure have actually been observed in related compounds of  the RMn$_2$O$_5$ family with R a rare-earth atom \cite{deutsch_pressure-induced_2018,peng_tuning_2019}. They crystallize in the same space 
group, forming pentagons of Mn$^{3+}$ and Mn$^{4+}$, where the Mn tetrahedra are connected by two oxygens instead of one as in Bi$_2$Fe$_4$O$_9$. No pressure-induced structural transition has been reported yet for most members of the family but they present magnetically induced ferroelectricity which is also modified under pressure. \textbf{Associated with} the expected change of the magnetic structure, one may thus wonder about the possibility of magnetically-induced ferroelectricity under pressure in Bi$_2$Fe$_4$O$_9$ as well.


\section{Conclusion}

In summary, we have reported the infrared spectrum of the pentagonal antiferromagnet Bi$_2$Fe$_4$O$_9$ as a function of temperature and pressure and identified strong signatures of magnetoelastic effects. One peculiar phonon around 225 cm$^{-1}$ presents an abnormal softening both in temperature and pressure. This mode involves atomic displacements of all Fe and O in each pentagonal planes; it also dynamically favors an increase of coordination of the tetrahedral irons, which is expected to be the driving mechanism responsible for the pressure induced structural transition earlier reported. This is expected to modify significantly the respective strength of the competing magnetic interactions in Bi$_2$Fe$_4$O$_9$ and we therefore anticipate important modifications of the magnetic structure at 
high pressure in this remarkable material.

\section*{ACKNOWLEDGMENTS}
M.V. gratefully acknowledge financial support and beamtime from synchrotron SOLEIL. We thank Region centre for financial support for the high pressure set-up. The authors would like to thank B. Baptiste of the DRX platform of IMPMC laboratory (Paris 6) for the orientation of the sample before the experiment, M. Lheronde of the IPANEMA-CNRS platform for the availability of their surface preparation laboratory and M. Iliev for the help in dynamical calculations.

\section{Appendix}
Bi$_2$Fe$_4$O$_9$ reflectivity spectra at different temperatures and pressures have been fitted using the Drude-Lorentz model with the REFIT software \cite{RefFit}. More details on the fits and the associated parameters are reported in this section.\\

The Drude-Lorentz model describes the optical response of an insulator with a set of harmonic oscillators for each phonon mode. The dielectric constant is given by
\begin{equation}
\epsilon (\omega)= \epsilon_\infty + \sum_k \frac{\Delta\epsilon_k\Omega^2_{k}}{\Omega^2_{k}-\omega^2 - i \gamma_k\omega}
\label{epsilon}
\end{equation}

$\epsilon_\infty$ is the so called "high-frequency dielectric constant" which represents the contribution of all oscillators at very high frequencies. The parameters  $\Omega_{k}$, $\Delta\epsilon_k$, and $\gamma_k$ are 
the resonance frequency, the oscillator strength and the linewidth, respectively of the k-th Lorentz oscillator. 

\begin{figure}
\centering
\includegraphics[width=0.9\columnwidth]{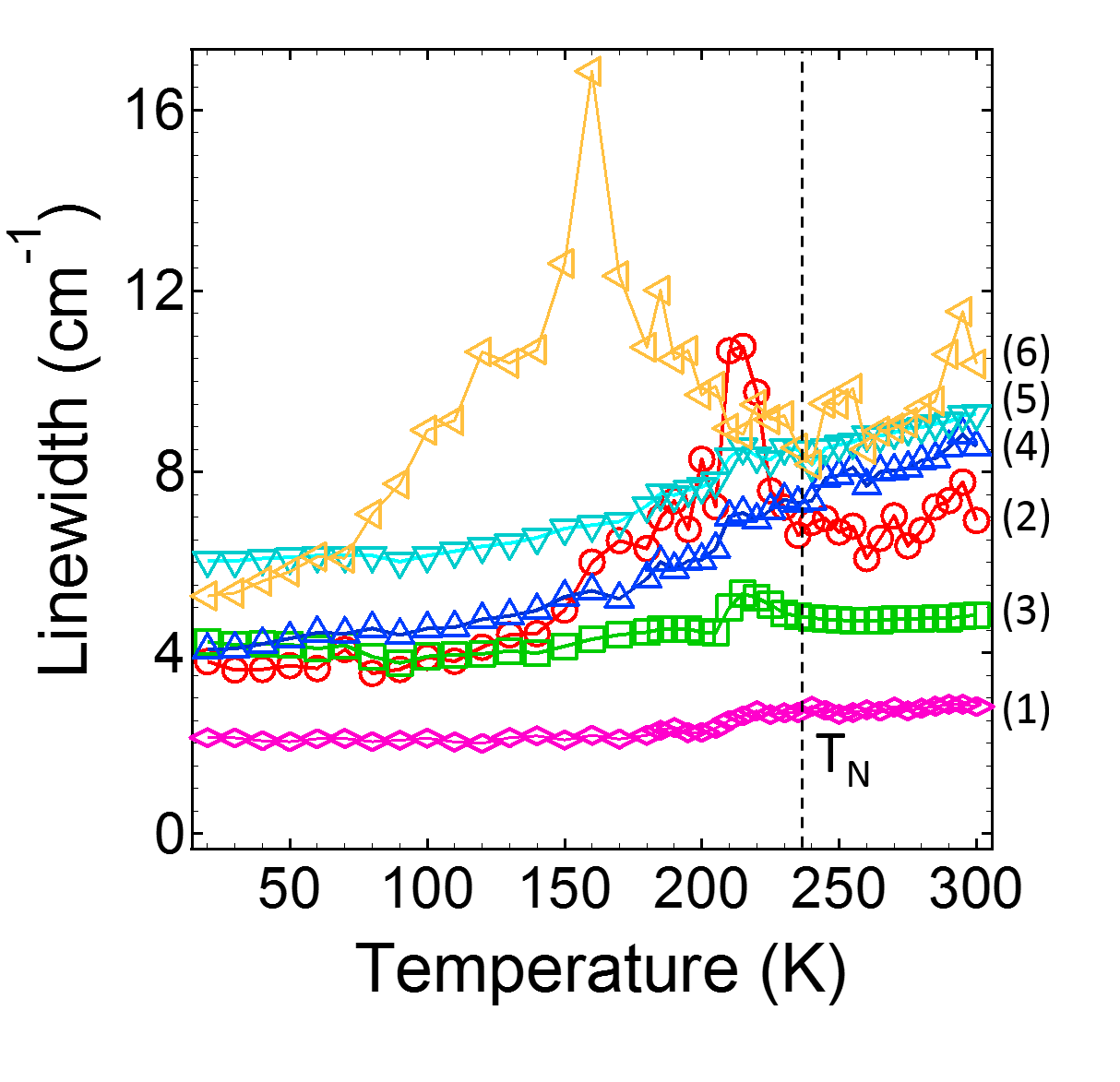}
\caption{Linewidth ($\gamma$) of modes B$_{2u}$(1), B$_{2u}$(2), B$_{2u}$(3), B$_{2u}$(4), B$_{2u}$(5), B$_{2u}$(6) as a function of temperature. The dashed line represents the Néel temperature of the sample, T$_N$ = 238 K. }
\label{Supinfo_Fig1}
\end{figure}

The DL fit parameters at selected temperatures and ambient pressure are given in Table \ref{Table1supinfo} for modes (2)-(6). In Figure \ref{Supinfo_Fig1}, their linewidth  is plotted as a function of temperature. Mode (6) is anomalously large from 200 K down to 60 K. It involves Fe1 as well 
as O1 and O3 atomic sites. Mode (2) presents a much sharper anomaly just below $T_N$: this mode involves only Fe1 and O1, which corresponds to the 
strongest antiferromagnetic bound. Modes (1),(3) - (5) have weaker temperature dependencies.

\begin{table*}[htbp!]
\center
\begin{tabular}{c|c|c|c|c|c|c|c|c|c|c|c|c|c|c|c|c|c|c}
\hline
\hline
 T(K)&\multicolumn{3}{c|}{Mode (1)}&\multicolumn{3}{c|}{Mode (2)} & \multicolumn{3}{c|}{Mode (3)}& \multicolumn{3}{c|}{Mode (4)} & \multicolumn{3}{c|}{Mode (5)} & \multicolumn{3}{c}{Mode (6)}\\
& $\Omega$  & $\gamma$  & $\Delta\epsilon$ & $\Omega$ & $\gamma$ & $\Delta\epsilon$ & $\Omega$ & $\gamma$  & $\Delta\epsilon$ & $\Omega$& $\gamma$& $\Delta\epsilon$ & $\Omega$& $\gamma$ &  $\Delta\epsilon$ \\

 & {\footnotesize (cm$^{-1}$)} & {\footnotesize (cm$^{-1}$)}&  &{\footnotesize (cm$^{-1}$)} & {\footnotesize (cm$^{-1}$)}&  & {\footnotesize (cm$^{-1}$)} & {\footnotesize (cm$^{-1}$)} &  & {\footnotesize (cm$^{-1}$)} & {\footnotesize (cm$^{-1}$)} & & {\footnotesize (cm$^{-1}$)}& {\footnotesize (cm$^{-1}$)}&  & {\footnotesize (cm$^{-1}$)} & {\footnotesize (cm$^{-1}$)} & \\
\hline
20 & 69.6 & 2.1 & 2.08 & 111.3 & 3.8 & 0.55 & 131.7 & 4.2 & 2.00 & 224.6 & 4.1 & 0.89 & 242.1 
& 6.0 & 0.79 & 311.3 & 5.3 & 0.35 \\
\hline
50 & 69.7 & 2.0& 2.01& 111.4 & 3.7  & 0.55 & 131.6 & 4.2 & 2.07 & 224.6 & 4.3 & 0.95 & 241.8 & 6.1 & 0.82 & 311.1 & 5.8 & 0.39  \\
\hline
100 & 69.7&2.1 &2.43 & 111.3 & 3.9 & 0.70 & 130.8 & 3.9 & 2.61 & 224.3 & 4.5  & 1.37 & 240.5 & 6.2 & 1.09 & 309.9  & 8.9  & 0.83     \\
\hline
150 &69.5 & 2.1&2.74 & 111.0 & 4.9 & 0.95 & 129.7 & 4.1 & 2.91 & 224.7 & 5.2 & 1.41 & 239.9 & 6.8 & 1.20 & 309.3 & 12.6 & 1.18    \\
\hline
200 &69.2 &2.2 &2.93 & 110.6  & 8.3 & 1.37 & 128.9 & 4.5 & 2.87 & 225.4 & 6.1  & 1.13 & 239.9 & 7.7 & 1.14 & 310.1 &  9.7 & 0.88     \\
\hline
220 & 69.2& 2.7&2.39 & 110.5 & 9.8 & 1.06 & 129.2 &5.2  & 2.29 & 226.0 & 6.9 & 0.73 & 240.7 & 8.3 & 0.82 & 310.8  & 9.5 &  0.36    \\
\hline
230 & 69.1& 2.7& 2.67& 110.6 & 7.2 & 0.85 & 129.3& 4.9 & 2.60 & 225.6 & 7.4 & 0.98 & 240.1 
& 8.5  & 0.95 & 310.8  & 9.2 & 0.42    \\
\hline
240 & 69.0& 2.8& 2.88& 110.6  & 6.9  & 0.82 & 129.2 & 4.8 & 2.71 & 225.6 & 7.4 & 1.00 & 240.0 & 8.2 & 0.98 & 310.8 & 8.2 &  0.37   \\
\hline
250 & 69.0&2.6 &2.81 & 110.6 & 6.7 & 0.82 &129.0  & 4.7 & 2.77  & 225.3 & 7.9 & 1.13 & 239.6  & 8.6 & 0.98 &310.6  & 9.5 &  0.44 \\
\hline
300 &68.9 & 2.8& 2.98& 110.5 &6.9 & 0.90 &128.3 &4.8 &3.08 &224.9 & 8.6 & 1.33 & 238.5 &9.3 & 1.06 & 310.1 &10.4 & 0.44 \\
\end{tabular}
\caption{Ambient pressure DL phonon parameters for modes (1)-(6) obtained 
from the fits of the reflectivity measured at selected temperatures.}
\label{Table1supinfo}

\end{table*}

\begin{figure}
\centering
\includegraphics[width=0.9\columnwidth]{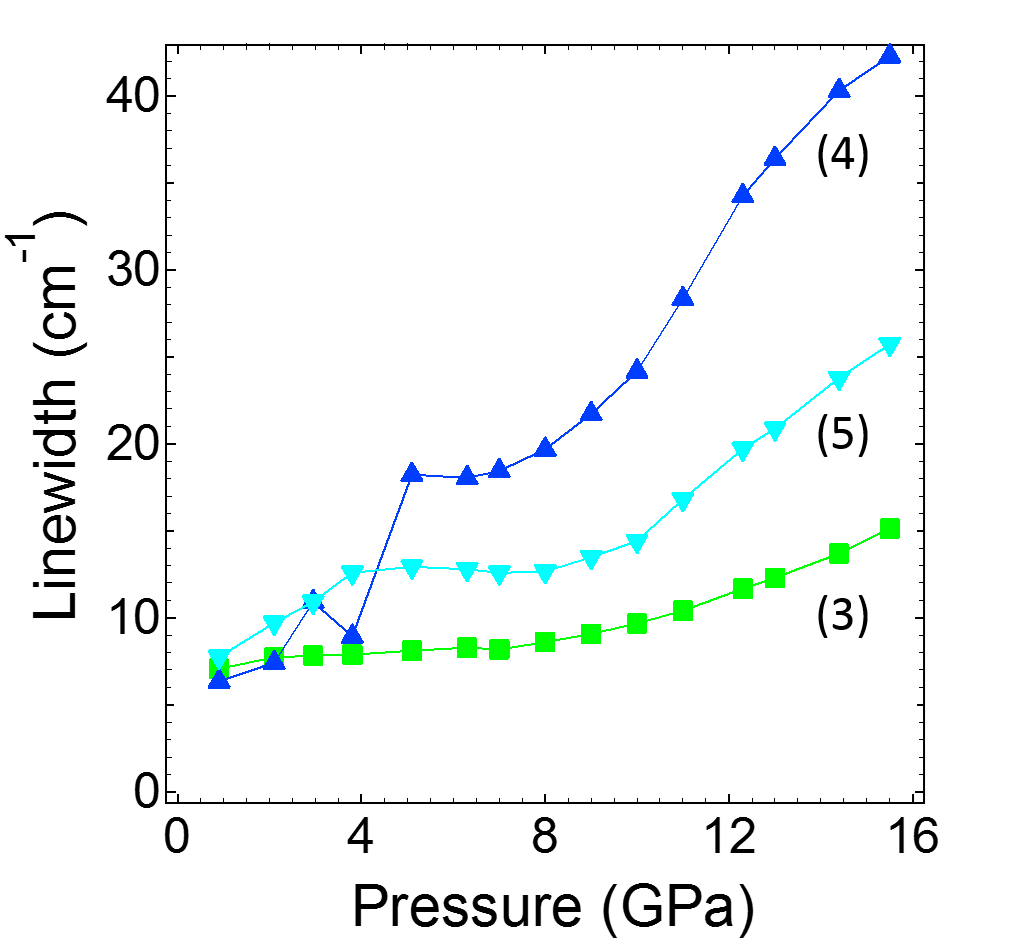}
\caption{Pressure dependence of linewidths ($\gamma$) of modes B$_{2u}$(3), B$_{2u}$(4)and B$_{2u}$(5) of Bi$_2$Fe$_4$O$_9$ obtained from the Drude-Lorentz fits of measurements performed at T = 40 K as a function of pressure.}
\label{Supinfo_Fig3}
\end{figure}

In Table \ref{Table2supinfo} are reported the DL phonon parameters obtained from the reflectivity measured at 40 K  and different pressures. The fits have been performed for modes (3) to (5). Their linewidth is plotted in Figure \ref{Supinfo_Fig3} as a function of pressure up to 15.5 GPa. No 
clear anomaly is observed but an overall tendency to an increase with increased pressure, with an upturn above 9 GPa.

\begin{table*}
\center
\begin{tabular}{c|c|c|c|c|c|c|c|c|c}
\hline
\hline
 P(GPa)& \multicolumn{3}{c|}{Mode (3)}& \multicolumn{3}{c|}{Mode (4)} & \multicolumn{3}{c}{Mode (5)} \\
& $\Omega$  & $\gamma$ & $\Delta\epsilon$ & $\Omega$ & $\gamma$ & $\Delta\epsilon$ & $\Omega$ & $\gamma$ & $\Delta\epsilon$ \\
 & {\footnotesize (cm$^{-1}$)} & {\footnotesize (cm$^{-1}$)}&  & {\footnotesize (cm$^{-1}$)} & {\footnotesize (cm$^{-1}$)} &  & {\footnotesize (cm$^{-1}$)} & {\footnotesize (cm$^{-1}$)} &  \\
\hline
\hline
 0.9 & 130.9 & 7.1 & 0.42 & 222.2 & 6.3 & 0.22 & 244.4 & 7.8 & 0.19  \\
\hline
 2.1 & 131.2 & 7.7 & 0.44 & 220.6 & 7.4 & 0.23 & 247.2 & 9.7& 0.18  \\
\hline
 3.0 & 131.5 & 7.8 & 0.48 & 219.5 & 10.9 & 0.26 & 248.9 & 10.9 & 0.20  \\
\hline
 3.8 & 131.9 & 7.9 & 0.49 & 217.9 & 8.9 & 0.28 & 251.2 & 12.6 & 0.22 \\
\hline
5.1 & 132.1 & 8.1 & 0.50 & 213.1 & 18.2 & 0.42 & 253.1 & 12.9 & 0.24   \\
\hline
6.3 & 132.3 & 8.3 & 0.52 & 212.1 & 18.0 & 0.42 & 254.3 & 12.8 & 0.25   \\
\hline
7   & 132.5 & 8.2 & 0.53 & 211.7 & 18.4 & 0.44 & 255.5 & 12.6& 0.25  \\
\hline
8   & 132.6 & 8.6 & 0.55 & 210.5  & 19.7 & 0.47 & 257.3 & 12.7 & 0.27  \\
\hline
9   & 132.6 & 9.1 & 0.55 & 209.6 & 21.7 & 0.49  & 258.5 & 13.5& 0.28   \\
\hline
10  & 132.8 & 9.7 & 0.56 & 208.6 & 24.2 & 0.51 & 259.6 & 14.4 & 0.29   \\
\hline
11  & 132.8 & 10.4  & 0.56 & 208.3 & 28.3 & 0.54 & 259.0 & 16.8& 0.29    \\
\hline
12.3 & 133.0 & 11.7 & 0.57 & 208.4 & 34.3 & 0.62 & 259.3  & 19.7& 0.29  \\
\hline
13 & 132.9 & 12.3  & 0.58 & 208.6 & 36.4 & 0.65 & 259.4 & 20.9 & 0.28   \\
\hline
14.4& 133.0 & 13.7  & 0.60  & 210.4 & 40.3 & 0.69 & 258.9 & 23.8& 0.28   \\
\hline
15.5& 132.6 & 15.1  & 0.65 & 212.3 & 42.3 & 0.66 & 259.6 & 25.7& 0.26 \\
\end{tabular}
\caption{Phonon parameters obtained from the fits of the reflectivity measured at 40 K for different value of pressure. }
\label{Table2supinfo}

\end{table*}


\bibliography{Bibliographie}

\end{document}